\documentclass[letterpaper,journal]{IEEEtran}
\usepackage{orcidlink}
\usepackage{cite}
\usepackage{amsmath,amssymb,amsfonts}
\usepackage{graphicx}
\usepackage{textcomp}
\usepackage{xcolor}
\usepackage[utf8]{inputenc}
\usepackage{algorithm}
\usepackage{algpseudocode}
\usepackage{comment}
\usepackage{graphicx} 
\usepackage{tabularx}
\usepackage{tikz}
\usepackage{booktabs} 
\usepackage{booktabs,multirow,pifont}
\usepackage{subcaption}
\usepackage[font=footnotesize,compatibility=false]{caption}

\usepackage{hyperref}

\begin{document}

\title{Fast Wireless Foundation Models with Early-Exits}
\author{Omar~Mashaal,~\IEEEmembership{Student~Member,~IEEE}, 
        and Hatem~Abou-Zeid,~\IEEEmembership{Member,~IEEE}%

\thanks{Manuscript received XX XX, 2025; revised XX XX, 2025; accepted XX XX, 2025.
Date of publication XX XX, 2025; date of current version XX XX, 2025. This work was supported in part by Alberta Innovates and in part by the Natural Sciences and Engineering Research Council of Canada (NSERC) under Grant RGPIN-2021-04050.
(Corresponding author: Hatem~Abou-Zeid.)}%
\thanks{The authors are with the Department of Electrical and Software Engineering,
University of Calgary, Calgary, AB T2N 1N4, Canada
(e-mail: omar.mashaal1@ucalgary.ca; hatem.abouzeid@ucalgary.ca).}%
}

\maketitle

\begin{abstract}

While wireless foundation models (FMs) are demonstrating strong potential to enable AI-Native 6G networks, their high computational cost remains a critical barrier to deployment. The large computational cost stems from the rigid, full-depth execution of the FM backbone for every task $-$ a process we show is not only inefficient but can also degrade performance on unseen out-of-distribution (OOD) tasks. In this paper, we propose a novel early-exit FM framework that attaches lightweight, per-task heads, at the most appropriate exit-stage of a frozen wireless FM encoder, enabling variable-depth inference tailored to each task’s preferred representation depth.
Our results demonstrate that these intermediate-layer features not only speed-up inference significantly (up to 93\% fewer FLOPs), but also provide more transferable representations that exceed the full encoder accuracy on unseen tasks. We further demonstrate that a simple fixed-exit strategy per task is more effective than traditional early-exiting policies that route different samples to different exits based on their perceived difficulty levels. 
\end{abstract}


\section{Introduction}

\IEEEPARstart{T}{he} vision for AI-native 6G wireless systems involves embedding intelligence to enhance network performance, spectrum efficiency, and adaptability \cite{brik2022deep}. Deep learning has already shown strong potential across many wireless applications, such as modulation recognition, channel estimation, and waveform design—typically through supervised learning (SL). While SL enabled early breakthroughs, it faces key limitations: labeled data is often scarce and expensive, models are task-specific, and they struggle to generalize in dynamic wireless environments. Even small shifts in conditions can require full retraining, making large-scale deployment impractical.

Wireless foundation models (WFMs) learn general representations for diverse tasks and environments. Recent works have explored WFMs for CSI, spectrograms~\cite{r_other_7, r_other_8, r_other_6}, and raw-IQ streams, including multi-task SSL~\cite{Kanu_IQ} and our prior work IQFM~\cite{IQFM}. However, their runtime cost and inference latency hinder deployment on resource-limited devices.

\begin{figure*}[!t]
    \centering
    \includegraphics[width=0.61\linewidth,height=0.21\linewidth]{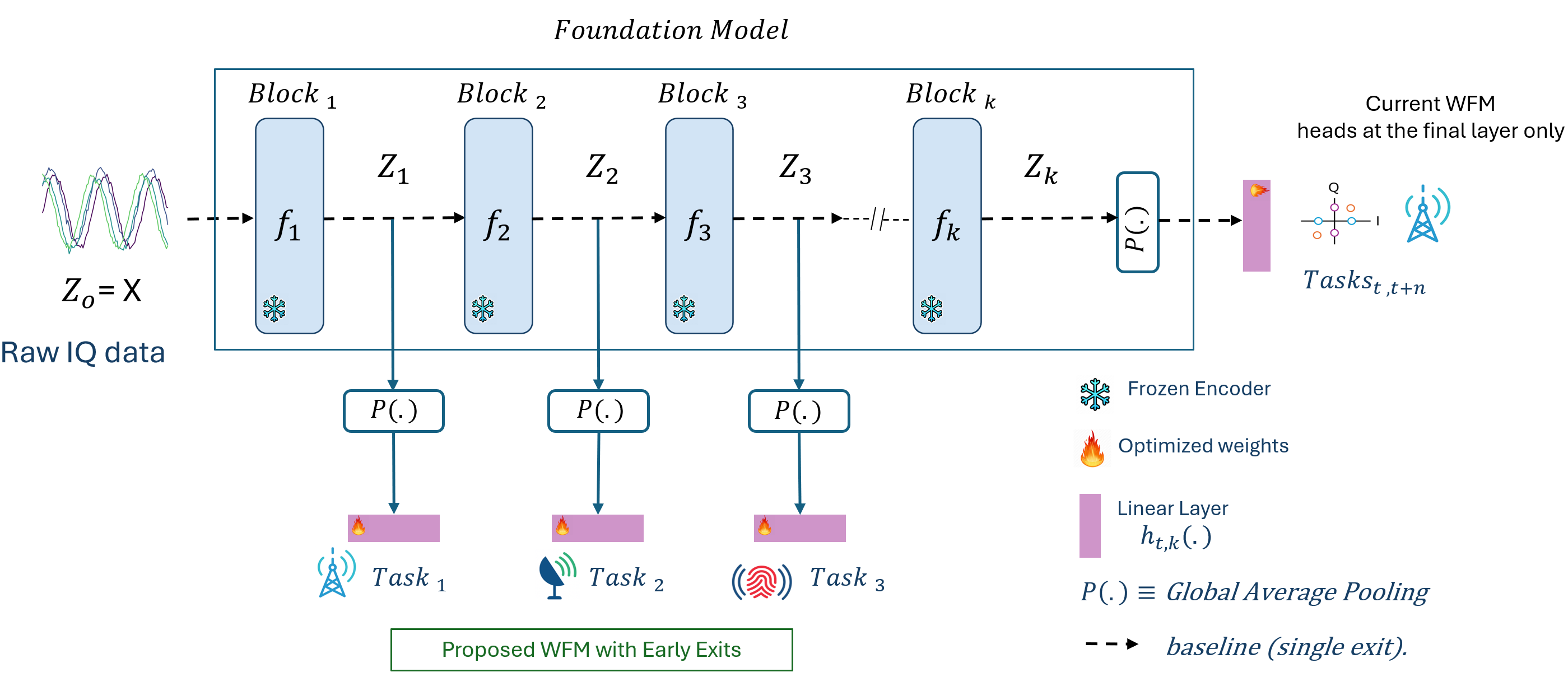}

\caption{Proposed Wireless Foundation Model with Early Exits.}

    \label{fig:iqfm_multi_exit}
\end{figure*}

The high computational cost of WFMs can be attributed to the design convention of exclusively using features from the encoder's final layer for downstream tasks. This practice requires executing the full backbone for every task, a process we demonstrate can be suboptimal. Our intuition is that features from shallower layers are often more generalizable and may be sufficient for general out-of-distribution tasks. In addition, the final layer's features carry the risk of being over-specialized and biased toward the pretraining data. By attaching lightweight heads to these intermediate layers (Fig.\ref{fig:iqfm_multi_exit}), we can bypass the full computational path and improve performance, especially for unseen, out-of-distribution (OOD) downstream tasks. This motivates our central question: is a full-depth execution of a WFM always necessary $-$ \textbf{or is it possible to identify the shallowest exit \emph{per task} that preserves or improves the performance while substantially reducing inference time?}

This notion of branching is known as early exiting (EE)~\cite{branchynet}, which reduces inference latency by allowing predictions at intermediate layers. In wireless, EE has been applied to single-task networks, such as for AMC~\cite{elsayed_ee_amc, widthwise_ee_amc} and collaborative edge-cloud inference~\cite{adaptive_early_exiting_collab}. These prior works typically allow easier samples (e.g., high-SNR signals) to exit early based on a confidence threshold.In contrast, we consider a frozen WFM adapted independently to multiple downstream tasks, where the dominant variation is across tasks rather than across samples within a single task. Different tasks may therefore prefer different representation depths. As shown in Fig.~\ref{fig:iqfm_multi_exit}, we attach lightweight task-specific heads at intermediate stages and compare classical confidence-based routing with a fixed-exit policy that selects one task-appropriate exit for each task. Unlike conventional EE, which is typically framed as an accuracy--latency tradeoff, we find that in the WFM setting, intermediate exits reduce latency while also improving task performance.
Our contributions are:
\begin{itemize}
\item We propose the first multi-exit neural architecture for wireless FMs. This architecture  attaches lightweight, per-task heads at the most appropriate encoder depths to enable task-specific inference paths that significantly reduce latency and computational cost.

\item We demonstrate that this approach reduces inference cost significantly (up to 93\% fewer FLOPs) while concurrently surpassing the full encoder accuracy on unseen tasks by up to 8.0\%, thereby showing that the typical accuracy-cost tradeoff of EE need not hold in this setting.

\item We provide a comprehensive, per-task analysis of our multi-exit WFM architecture across diverse wireless tasks including RF fingerprinting, beam prediction, and classification of interference signals and modulation types. 

\item  We propose and evaluate a few-shot best-exit selection method that identifies the best exit for each task. We compare this strategy against classical dynamic-routing policies, including \emph{Pareto} and \emph{Greedy} EE approaches. Our results show that 
the fixed-exit captures all, or most, of the achievable benefit while avoiding the additional dynamic routing
and threshold-checking overhead.
\end{itemize}

By converting a single-path WFM into a multi-exit pipeline, our approach delivers predictable low latency and substantial compute and latency savings, while improving performance on new tasks. This paper provides the foundation and demonstrates the strong potential of such EE architectures for WFMs.

\section{Multi-Exit Wireless Foundation Model}
\subsection{Signal and Input Representation}
We consider narrowband, far-field reception at a uniform linear array (ULA) with \(M\) antennas. The complex baseband signal at antenna \(m\) is
\begin{equation}
x_m(t)=\alpha\,s(t)\,e^{j\frac{2\pi}{\lambda}p_m\sin\theta}+n_m(t),\quad
p_m=(m{-}1)\,d_{\mathrm{ant}},
\label{eq:rx}
\end{equation}

\noindent where \(s(t)\) is the transmitted baseband waveform, \(\alpha\) is the complex path gain, \(\lambda\) the wavelength, \(d_{\mathrm{ant}}\) the inter-element spacing, and \(\theta\) the angle of arrival. The additive noise \(n_m(t)\) is typically modeled as AWGN, though in over-the-air datasets it may also encompass practical impairments. The received signal is then sampled and stacked into an IQ tensor \(X\in\mathbb{R}^{M\times 2\times T}\), where $M$ is the number of antennas and $T$ is the number of time samples. This format preserves both array structure and temporal context for downstream learning tasks.

\begin{algorithm}[tb]
\caption{Training Early-Exit Task Heads}
\label{alg:train_heads}
\begin{algorithmic}[1]
\State \textbf{Input:} frozen encoder blocks $f_{1:K}$; pooling $P(\cdot)$; task dataset $\mathcal{D}_t=\{(X,Y)\}$
\State \textbf{Output:} trained heads $\{h_{t,k}\}_{k=1}^K$
\For{$k=1$ to $K$}
  \State Initialize head $h_{t,k}$ \Comment{same arch/hparams for all $k$}
  \For{epochs}
    \For{mini-batch $(X,Y)\subset\mathcal{D}_t$}
      \State $Z_k \gets f_k(\cdots f_1(X))$ \Comment{encoder frozen}
      \State $\mathbf{u}_k \gets P(Z_k)$ \Comment{global avg pooling}
      \State $\widehat{y}^{[k]}_t \gets h_{t,k}(\mathbf{u}_k)$
      \State $\mathcal{L}\gets$ CE for classification or MAE for regress.
      \State update parameters of $h_{t,k}$ only
    \EndFor
  \EndFor
\EndFor
\end{algorithmic}
\end{algorithm}

\subsection{Proposed Foundation Model with Multiple Exits}
As shown in Fig.~\ref{fig:iqfm_multi_exit}, we use a pretrained raw IQ foundation encoder presented in \cite{IQFM} as a frozen backbone decomposed into \(K\) sequential blocks:
\begin{equation}
Z_0 = X,\qquad Z_k = f_k(Z_{k-1}),\; k=1,\ldots,K,
\label{eq:blocks}
\end{equation}
The intermediate feature set after block \(k\) (Stage \(k\)) is \(Z_k\). Each stage defines an early exit; we apply global average pooling (GAP) \(P(\cdot)\) to obtain \(\mathbf{u}_k = P(Z_k)\), and attach a lightweight head \(h_{t,k}\) for task \(t\),

\begin{equation}
\widehat{y}_t^{[k]} = h_{t,k}(\mathbf{u}_k),
\label{eq:head}
\end{equation}

We index the frozen backbone blocks $\{f_k\}$ by $k$, where $K=5$ corresponds to the major computational blocks of the ShuffleNetV2-x0.5 stages. This structure allows the WFM to exit at task-specific depths rather than a fixed final point, enabling improved performance and efficiency.

\section{Training and Exit Selection Methodology}

We adapt a frozen IQ foundation model by training lightweight heads at multiple exit points. Unlike classical early-exiting methods, our approach performs a one-time, offline selection of the optimal inference depth for each task. We first describe the training of the early-exit heads, followed by the few-shot and unlabeled selection strategies.

\subsection{IQFM Pretraining using Contrastive Learning}

We adopt the IQFM backbone pretrained via self-supervised SimCLR/InfoNCE learning on raw multi-antenna IQ data~\cite{IQFM}.

\subsection{Head Architecture and Training Procedure}

We train heads at all exits on the frozen encoder and select the top-performing exit per task. This procedure, which yields one specialized head per task and per exit stage, is formalized in Algorithm~\ref{alg:train_heads}. For each task $t$ and exit stage $k$, the corresponding feature map $\mathbf{Z}_k$ is first globally averaged to a vector $\mathbf{u}_k = \operatorname{\mathbf{P}}(\mathbf{Z}_k)$. This vector is then processed by a linear layer to produce logits,
\begin{align}
\mathbf{z}_{t,k} &= \mathbf{W}_{t,k}\,\mathbf{u}_k + \mathbf{b}_{t,k}, \\
\mathbf{p}_{t,k} &= \operatorname{softmax}(\mathbf{z}_{t,k}).
\end{align}

where $\mathbf{z}_{t,k}$ represents the logits for exit $\mathbf{k}$ and task $\mathbf{t}$, i.e., the unnormalized scores generated by the final linear layer, and $\mathbf{p}_{t,k}$ denotes the predicted class probabilities. To ensure a fair comparison that isolates the quality of the encoder's representations rather than the head's capacity, all heads share an identical architecture and are trained with the same hyperparameters (e.g., AdamW optimizer, learning-rate schedule). In addition, each exit uses GAP to form a compact fixed-size representation before classification. Our experiments with more expressive heads showed the same overall preferred-exit trend, indicating that the main conclusion is not specific to the linear head design. Each head is trained independently to minimize the standard cross-entropy loss on its task-specific training data:

\begin{equation}
\mathcal{L}^{\mathrm{cls}}_{t,k}
= -\frac{1}{N}\sum_{i=1}^{N} \log p^{(i)}_{t,k}\!\left(y_i\right).
\end{equation}

\begin{algorithm}[t]
\caption{Classical Early-Exit Inference (adapted from prior 
confidence-based early-exit methods~\cite{branchynet})}
\label{alg:ee_infer}
\begin{algorithmic}[1]
\State \textbf{Input:} IQ input $x$; frozen encoder $f_{1:K}$; 
  pooling $P$; heads $\{h_{t,k}\}_{k=1}^K$; normalized entropy 
  thresholds $\{T_k\}_{k=1}^{K-1}$, with $T_k = \bar{\tau}\ln C_t$ 
  for all $k < K$, $\bar{\tau}\in[0,1]$
\State \textbf{Output:} predicted label $\hat{y}$
\State $Z_0 \leftarrow x$
\For{$k=1$ \textbf{to} $K$}
    \State $Z_k \leftarrow f_k(Z_{k-1})$
    \State $\mathbf{u}_k \leftarrow P(Z_k)$;\; 
      $p_{t,k} \leftarrow \mathrm{softmax}\!\big(h_{t,k}(\mathbf{u}_k)\big)$
    \State $H_k \leftarrow -\sum_c p_{t,k}(c)\log p_{t,k}(c)$ 
      \Comment{predictive entropy}
    \If{$k < K$ \textbf{and} $H_k \le T_k$} 
      \Comment{confidence test}
        \State \Return $\arg\max_c\, p_{t,k}(c)$
    \EndIf
\EndFor
\State \Return $\arg\max_c\, p_{t,K}(c)$
\end{algorithmic}
\end{algorithm}

\begin{algorithm}[t]
\caption{Proposed Exit Selection and Inference Algorithm}
\label{alg:selected_infer}
\begin{algorithmic}[1]
\State \textbf{Input:} IQ inputs $\mathcal{X}$; frozen encoder blocks 
  $f_{1:K}$; pooling $P$; heads $\{h_{t,k}\}_{k=1}^K$; few-shot set 
  $\mathcal{S}$ or unlabeled set $\mathcal{U}$
\State \textbf{Output:} predicted labels $\hat{\mathcal{Y}}$; chosen 
  exit $k_t^\star$
\State \textbf{One-time exit selection (offline, per task):}
\If{$\mathcal{S}$ available} \Comment{few-shot; }
    \State $k_t^\star \leftarrow \arg\max_k\, \widehat{A}_t(k;\mathcal{S})$
      \Comment{few-shot val.\ acc.}
\Else \Comment{unlabeled}
    \State $\bar{c}_k \leftarrow \dfrac{1}{|\mathcal{U}|}
      \sum_{x\in\mathcal{U}} \max_c\, p_{t,k}(c\mid x)$
      \quad for each $k$
    \State $k_t^\star \leftarrow \arg\max_k\, \bar{c}_k$
      \Comment{equiv.\ $\arg\min_k\, \bar{H}_k$;}
\EndIf
\State $\hat{\mathcal{Y}} \leftarrow [\ ]$
\For{\textbf{each} $x$ \textbf{in} $\mathcal{X}$}
  \Comment{fixed-depth inference at $k_t^\star$}
    \State $Z_0 \leftarrow x$;\;
      \textbf{for } $k{=}1..k_t^\star$: $Z_k \leftarrow f_k(Z_{k-1})$
    \State $\mathbf{u} \leftarrow P(Z_{k_t^\star})$;\;
      $p \leftarrow \mathrm{softmax}\!\big(h_{t,k_t^\star}(\mathbf{u})\big)$
    \State append$\!\left(\hat{\mathcal{Y}},\;
      \arg\max_c\, p(c)\right)$
\EndFor
\State \Return $\hat{\mathcal{Y}},\, k_t^\star$
\end{algorithmic}
\end{algorithm}

\subsection{Inference Routing and Exit Selection Strategies}\label{sec:inference_routing}

We consider two deployment profiles that differ only in how an input is routed through exits.

\textbf{Classical Early Exits:}
At inference, after stage $k$, the head produces probabilities $p_{t,k}(\cdot\mid x)$. If the predictive entropy $H(p_{t,k}) \le T_k$ (where lower entropy implies higher confidence), the model exits; otherwise, it proceeds to stage $k{+}1$ (see Algorithm~\ref{alg:ee_infer}). To make thresholds comparable across tasks, we normalize them by the maximum entropy of the task. Specifically, we set $T_k=\bar{\tau}\ln C_t$, where $\bar{\tau}\in[0,1]$ is a normalized threshold and $\ln C_t$ is the maximum entropy for a $C_t$-class task. While dynamic EE allows easy inputs to exit early, it introduces input-dependent runtime, can exit at suboptimal depths, and its gating logic (entropy computation and branching) can add latency over a single fixed exit.

\textbf{Proposed Single Exit Selection Algorithm:}
Under the fixed-exit policy, we choose a single exit $k_t^\star$ per task once, offline, and then run all inputs to that fixed depth (Algorithm~\ref{alg:selected_infer}). Since FMs are designed to adapt with few labels, this selector is designed to operate with minimal or no labels. The exit is determined either from:

\emph{1) Unlabeled data exit selection.}
For each exit $k$, define probabilities by
$p_{t,k}(\cdot\mid x)=\mathrm{softmax}\!\big(h_{t,k}(P(f_{1:k}(x)))\big)$.
Per-sample confidence is $c_k(x)=\max_{c} p_{t,k}(c\mid x)$ and predictive entropy is
\[
H\!\left(p_{t,k}(\cdot\mid x)\right) = -\sum_{c=1}^{C_t} p_{t,k}(c\mid x)\,\log p_{t,k}(c\mid x)
\]
(in nats). Average over $\mathcal{U}$:
\[
\bar c_k = \tfrac{1}{|\mathcal{U}|}\sum_{x\in\mathcal{U}} c_k(x), \qquad
\bar H_k = \tfrac{1}{|\mathcal{U}|}\sum_{x\in\mathcal{U}} H\!\left(p_{t,k}(\cdot\mid x)\right).
\]
Select $k_t^\star=\arg\max_k \bar c_k$ (equivalently, $k_t^\star=\arg\min_k \bar H_k$), breaking ties toward smaller $k$.

\emph{2) Few-shot exit selection.}
Given $\mathcal{S}=\{(x_i,y_i)\}$, select
$k_t^\star=\arg\max_k\, \widehat{A}_t(k;\mathcal{S})$, where
\[
\widehat{A}_t(k;\mathcal{S})=
\frac{1}{|\mathcal{S}|}
\sum_{(x_i,y_i)\in\mathcal{S}}
\mathbf{1}\!\left\{\arg\max_c\, p_{t,k}(c\mid x_i)=y_i\right\},
\]
breaking ties toward smaller $k$.

\subsection{Datasets}
\label{sec:datasets}

For downstream evaluation, we use five datasets, categorized as either in-distribution (ID) or out-of-distribution (OOD) relative to the IQFM pretraining/adaptation setting. ID refers to the same data and task setting as in IQFM, whereas OOD refers to a shift in data distribution, task definition, or both.

\noindent\textbf{ID tasks:} We evaluate on Angle of Arrival (AoA) and modulation classification using the same in-house IQFM testbed data as in~\cite{IQFM}. These tasks represent the in-distribution setting. Collection details, including hardware setup, signal types, and the 225 AoA classes, are provided in~\cite{IQFM}.

\noindent\textbf{OOD tasks:} We evaluate on four downstream tasks using raw IQ inputs; all samples are zero-padded to match the model's 4-channel input.

\textit{DeepBeam (beam selection, task and distribution shift).}~\cite{polese2021deepbeam}
mmWave beam prediction with five discrete beams using provided beamformed IQ, from which we extract $(1,2,256)$ slices for 5-way classification.

\textit{RML2016.10a (modulation classification, distribution shift).}~\cite{Oshea_RML2016}
Eleven-class modulation recognition across SNRs from $-20$ to $+18$~dB, evaluated using standard $(1,2,128)$ IQ windows.

\textit{POWDER RF fingerprinting (device ID, task and distribution shift).}~\cite{reus2020trust}
Device identification from over-the-air Wi-Fi captures, using $(1,2,256)$ IQ samples for 4-way classification.

\textit{OWL-INT (interference classification, task and distribution shift).}~\cite{owlint2022}
Interference classification over Bluetooth/WiFi/802.15.4 IQ traces with 21 SNRs and channel offsets, using $(1,2,128)$ windows for 15-class classification.

\section{Results and Discussion}

We validate our framework on a frozen IQFM~\cite{IQFM} (ShuffleNetV2-x0.5) instrumented with exits after its computational blocks: Stage 2 (\textbf{S2}), Stage 3 (\textbf{S3}), and Stage 4 (\textbf{S4}), as shown in Fig.\ref{fig:iqfm_multi_exit}. We first compare each exit’s performance against the conventional final head (\textbf{Full}) on six evaluation tasks: two ID tasks and four OOD downstream tasks. We then study the classical EE accuracy-latency trade-off and evaluate our proposed unlabeled and few-shot selection methods. For each task/exit, we train a linear head on 500 samples/class for 100 epochs (AdamW) while keeping the encoder frozen. 
All experiments used an NVIDIA GeForce RTX 3080 Ti GPU.

\begin{table}[t]
\centering
\caption{Test accuracy (\%) at each exit per task (500 shots/class). 
\emph{Generalization:} ID = in-distribution w.r.t. IQFM pretraining; 
OOD = unseen datasets and/or tasks. Best per row in \textbf{bold}.}
\label{tab:acc_per_exit}

\setlength{\tabcolsep}{4pt}
\renewcommand{\arraystretch}{1.1}
\begin{tabular}{l l cccc}
\toprule
\textbf{Task} & \textbf{Generalization} & \textbf{S2} & \textbf{S3} & \textbf{S4} & \textbf{Full} \\
\midrule
RF fing. (RF\_ID).    & OOD & \textbf{90.461} & 86.830 & 86.016 & 83.166 \\

RML       & OOD & 63.774 & \textbf{71.783} & 69.755 & 65.866 \\
OWL-INT  & OOD &  \textbf{93.146} & 92.951 & 92.286 & 90.286 \\

DeepBeam  & OOD & 48.562 & \textbf{50.897} & 46.748 & 42.917 \\

AoA       &  ID & 97.280 & 99.419 & 99.750 & \textbf{99.754} \\
Mod       &  ID & 95.744 & 99.936 & 99.964 & \textbf{99.967} \\
\bottomrule
\end{tabular}
\end{table}

\subsection{Early Exits  Deliver Performance Gains and Efficiency}

\begin{figure}[t]
    \centering
    \begin{subfigure}[b]{0.495\columnwidth}
        \centering
        \includegraphics[width=\linewidth]{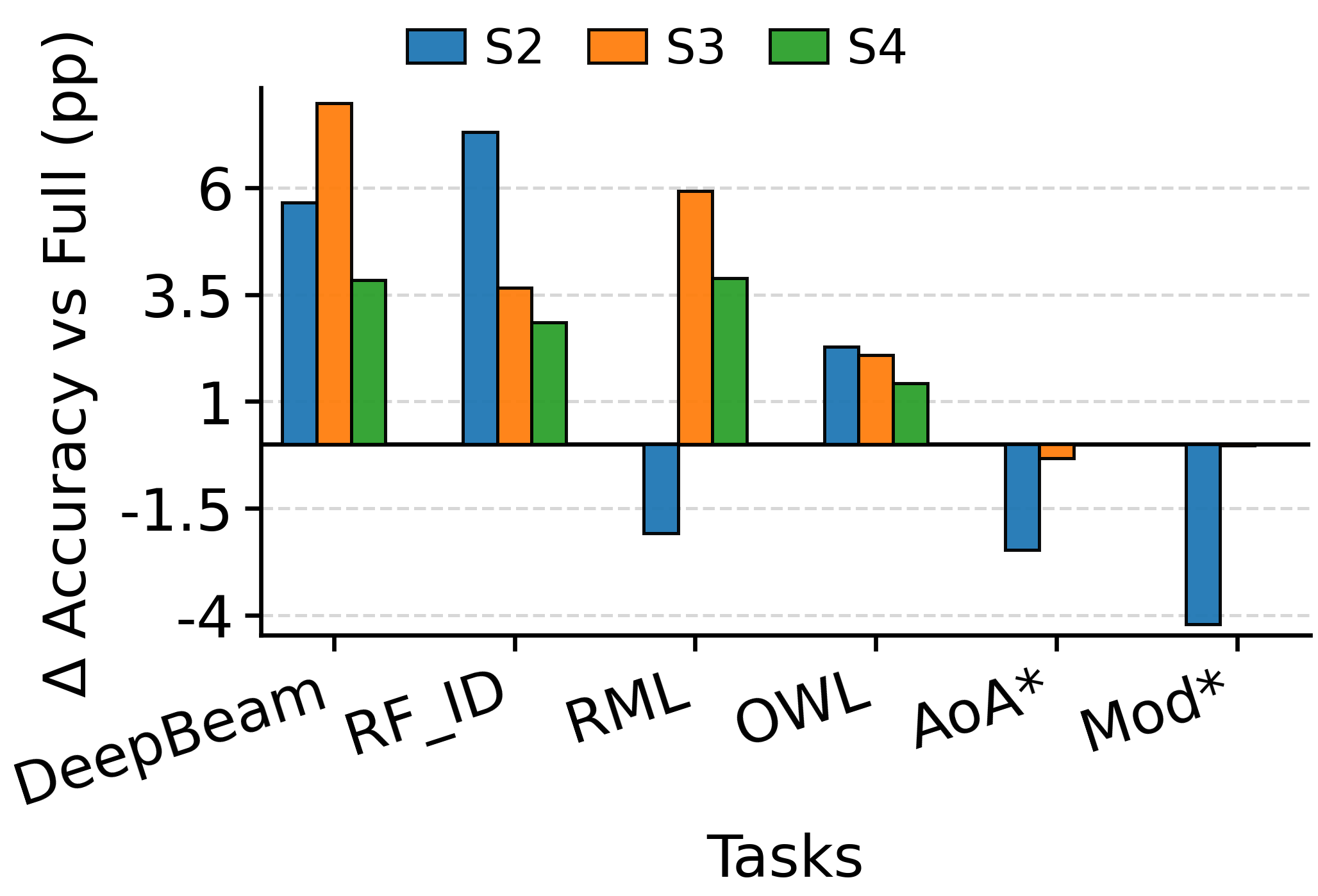}
        \caption{}
        \label{fig:combined_acc_ch_a}
    \end{subfigure}
    \hspace{0.00001\columnwidth}
    \begin{subfigure}[b]{0.48\columnwidth}
        \centering
        \raisebox{0.5cm}{\includegraphics[width=0.86\linewidth]{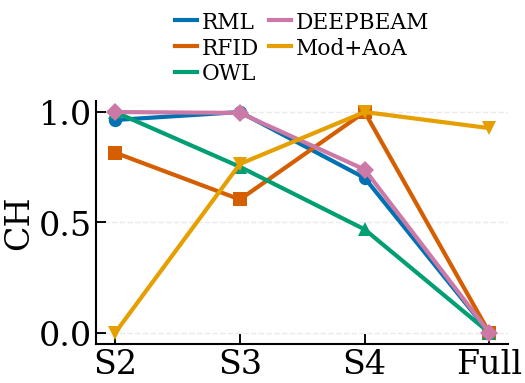}}
        \caption{}
        \label{fig:combined_acc_ch_b}
    \end{subfigure}
    \caption{(a) Test accuracy change (pp) of early exits (S2--S4) relative to the Full backbone. (b) Normalized CH trend across representative OOD tasks and the ID tasks.}
    \label{fig:combined_acc_ch}
\end{figure}

Our results demonstrate that for a frozen foundation model, shallower exits can deliver both performance gains and significant efficiency. As shown in Figure \ref{fig:combined_acc_ch_a} and Table \ref{tab:acc_per_exit}, intermediate exits outperform the full model on all four OOD tasks. Notably, S2 improves RF fingerprinting by +7.3 percentage points (pp). The S3 exit improves DeepBeam by +8.0 pp, OWL-INT by +2.3 pp, and RML (by +5.9 pp for SNRs $>$ 0, and from ~40.5\% to 43.2\% for all SNRs). 
 
For ID tasks, the Full exit remains optimal, but the S3 and S4 exits perform negligibly worse (within 0.335 pp). This indicates that the deepest features are not always optimal and can even be detrimental to generalization (Table~\ref{tab:acc_per_exit}).

These accuracy gains are paired with significant computational savings (Table \ref{tab:shortpathparameters}). Exiting at S2 reduces FLOPs by 93\% (a 5.6$\times$ speedup), while the S3 exit, which excels on three OOD tasks, has 71\% fewer FLOPs. These results suggest that the FM's final layers can over-specialize to the pretraining distribution, harming OOD, while intermediate representations transfer more robustly \cite{UselisOhICLR2025,LayerByLayerICML2025}. Fig.~\ref{fig:combined_acc_ch_b} supports these trends through normalized Calinski--Harabasz (CH) scores that quantify between-cluster separation relative to within-cluster spread. As shown, OOD tasks benefit from intermediate exits, whereas ID tasks have higher CH scores at deeper stages. Note that for RFID, the highest accuracy is at S2 but the CH is slightly higher at S4. PCA analysis indicates that RFID forms very dense clusters at S4 which increases the CH score. However, the overall cluster structure does not translate into better linear separability, resulting in lower accuracy compared with S2.

\begin{figure}[!t]
    \centering
    \includegraphics[width=0.85\linewidth]{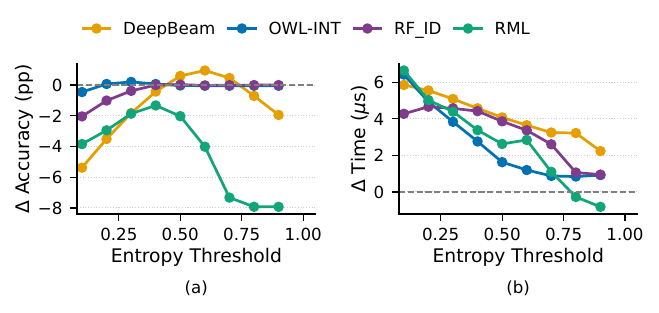}

   \caption{Performance of shared-entropy EE vs. best fixed exit. (a) Accuracy gain (higher is better); (b) Latency reduction (lower is better).}
  \label{fig:acc_lat}
\end{figure}

\begin{table}[!b]
    \centering
    \caption{\textbf{FLOPs and Number of Parameters cost by exit (cumulative up to the exit; heads excluded)}}
    \label{tab:shortpathparameters}
    \renewcommand{\arraystretch}{1.2}
    \setlength{\tabcolsep}{6pt}
    \begin{tabular}{lccc}
        \toprule
        \textbf{Model} & \textbf{FLOPs (M)} & \textbf{\# of Parameters} & \textbf{Avg. Infer. Time(us)}  \\
        \midrule
        Stage2       & 0.438      & 7,800   &  0.928\\
        Stage3        & 1.889     & 53,352  &  2.746 \\
        Stage4       & 3.433    & 143,304  &   4.286\\
        Full        & 6.579    & 341,960    &   5.238\\
        \bottomrule
    \end{tabular}
\end{table}

\begin{figure*}[!t]
\centering
\begin{subfigure}[t]{0.18\textwidth}\centering
  \includegraphics[width=\linewidth]{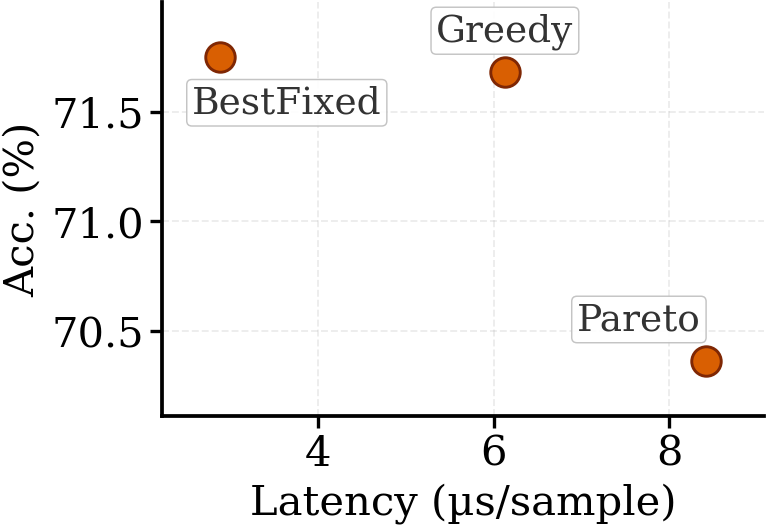}
  \caption{RML16: Acc. vs. latency}\label{fig:rml_acc_latency}
\end{subfigure}
\begin{subfigure}[t]{0.22\textwidth}\centering
  \includegraphics[width=\linewidth]{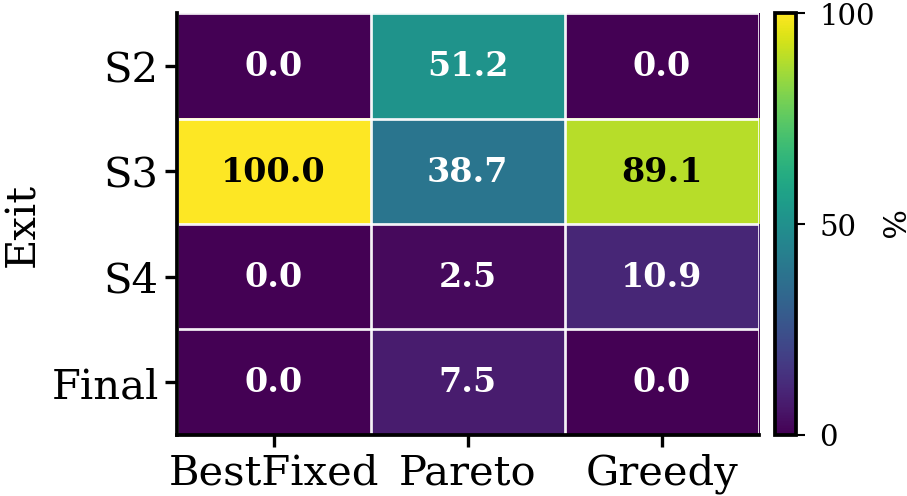}
  \caption{RML16: Exit-rate heatmap}\label{fig:rml_heatmap}
\end{subfigure}\hspace{0.07\textwidth}
\begin{subfigure}[t]{0.18\textwidth}\centering
  \includegraphics[width=\linewidth]{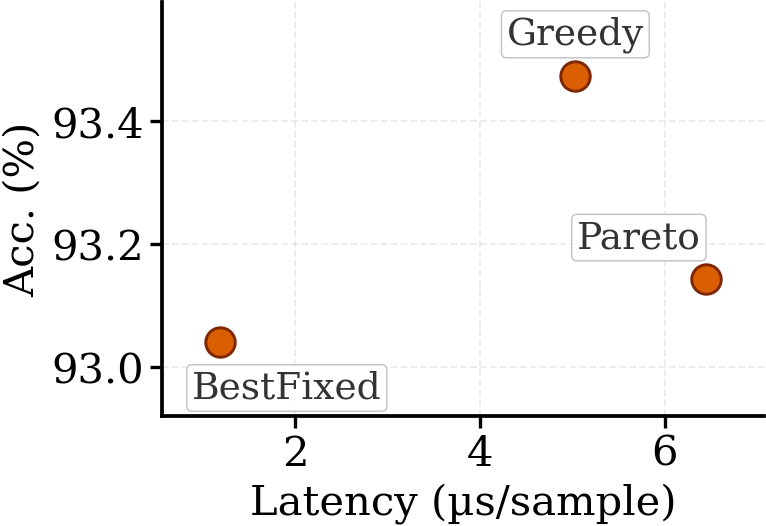}
  \caption{OWL: Acc. vs. latency}\label{fig:deepbeam_acc_latency}
\end{subfigure}
\begin{subfigure}[t]{0.22\textwidth}\centering
  \includegraphics[width=\linewidth]{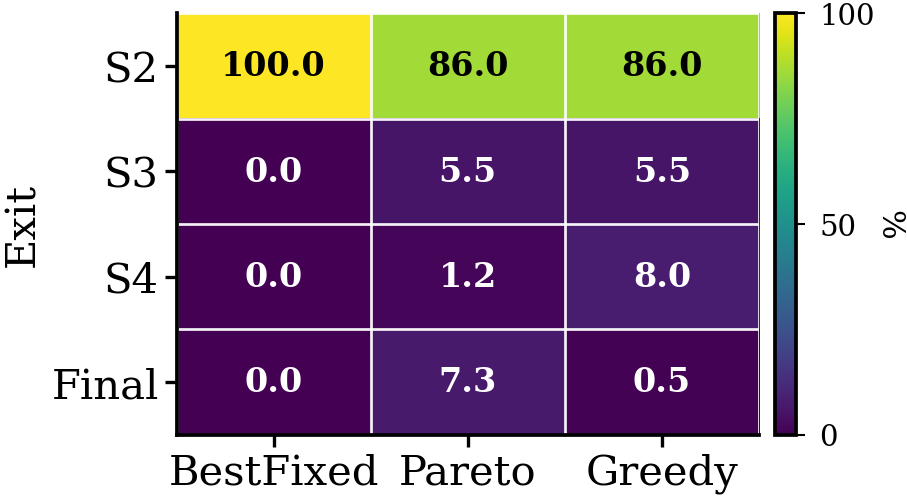}
  \caption{OWL: Exit-rate heatmap}\label{fig:deepbeam_heatmap}
\end{subfigure}
\caption{Representative dynamic-routing results on RML16 and OWL. Accuracy--latency plots compare BestFixed, Pareto, and Greedy; heatmaps show exit-rate allocations. Most samples concentrate at one dominant intermediate exit, explaining the competitiveness of BestFixed.}
\label{fig:dynamic_routing_examples}
\end{figure*}

\subsection{Dynamic Routing vs. Fixed-Exit Inference}

We next evaluate a classical EE policy (Algorithm~\ref{alg:ee_infer}) against our proposed fixed-exit-per-task strategy. We first study the shared-threshold setting by evaluating various entropy thresholds ($\tau$) (Sec.~\ref{sec:inference_routing}). Figure~\ref{fig:acc_lat}(a-b) plots the accuracy and latency ($\mu$s) change relative to the best fixed exit for that task, both as a function of the shared threshold $\tau$. These results show that the gains are generally small and inconsistent. We therefore study two stronger dynamic-routing variants as well. 
\emph{Greedy} is a per-exit tuned dynamic method that selects thresholds sequentially for the intermediate exits, allowing different exits to use different confidence thresholds.
\emph{Pareto} selects a shared threshold using validation data to determine the best  $\bar{\tau}$. We sweep $\bar{\tau}$ while computing validation accuracy and exit cost, and select the operating point with the highest validation accuracy, using lower cost as a tie-breaker.

Our results showed that these dynamic routing methods also offer no clear advantage over the fixed-exit baseline in the accuracy--latency tradeoff.
Representative results are shown in Fig.~\ref{fig:dynamic_routing_examples}. Here, \emph{BestFixed} serves as a fixed-exit reference, since our goal is to test whether dynamic routing provides a meaningful advantage beyond a carefully chosen fixed-exit policy.
For RML16, Greedy achieves only a slight accuracy gain over BestFixed, but at substantially higher latency, whereas Pareto is both slower and less accurate. A similar trend appears for OWL, where Greedy provides only a marginal improvement and Pareto again underperforms the fixed-exit baseline. The heatmaps help explain this behavior: in both tasks, routing remains heavily concentrated at one dominant intermediate exit, with limited benefit from distributing samples across multiple depths. This suggests that, in the considered WFM setting, a carefully selected fixed exit captures most of the achievable benefit while avoiding the additional routing and threshold-check overhead of dynamic methods.

\subsection{Choosing the Best Single Exit with Minimal Labels}\label{sec:exit_selection}

Having established the superiority of a single-exit approach, the remaining question is how to pick the exit while staying data-efficient. As detailed in Sec.~\ref{sec:inference_routing}, we evaluate three low-cost selectors: two label-free (average confidence, entropy) and a few-shot (FS) validator (\(k{=}20\)/class). Table~\ref{tab:best-exit} shows the label-free options can be misleading, confidence and entropy do not reliably track correctness, whereas few-shot validation consistently selects the best exit. We note that the required number of labeled samples is not universal and may depend on the dataset, sample representativeness, and the performance differences between candidate exits; a detailed analysis of this dependence is left for future work.


\begin{table}[!tpb]
\centering
\caption{Best-exit comparison across selection methods.}
\label{tab:best-exit}
\footnotesize
\setlength{\tabcolsep}{3pt}
\renewcommand{\arraystretch}{1}
\begin{tabular}{l c c c c}
\toprule
\textbf{Task} & \textbf{Best Exit} & \textbf{Confidence} & \textbf{Entropy} & \textbf{Few-shot} \\
\midrule
RML16      & \textbf{S3} & \textbf{S3}
                     & \textbf{S3} 
                     & \textbf{S3} \\
RF\_ID     & \textbf{S2} & S3
                     & S3 
                     & \textbf{S2} \\
DeepBeam   & \textbf{S3} & \textbf{S3}
                     & \textbf{S3}
                     & \textbf{S3}\\
OWL-INT   & \textbf{S2} & S3
                     & S3
                     & \textbf{S2}\\                     
\bottomrule
\end{tabular}
\end{table}

\section{Conclusion \& Future Work}

In this letter, we proposed a depth-aware framework that improves the inference speed of wireless foundation models. Our analysis reveals that exiting early from the encoder is not merely an efficiency trade-off but a key strategy for improving OOD generalization. Experimental results demonstrate that intermediate features provide noticeable accuracy gains (up to 8.0 pp) while drastically reducing latency. Furthermore, we showed that selecting a single, optimal exit for each task is a more robust strategy than complex dynamic multi-exit policies. The early-exit framework is not tied to a specific backbone and can be extended to other architectures. Future work will focus on developing more label-efficient and ideally fully label-free, exit selection methods and studying exit stability under large post-deployment channel and hardware variations.
Further optimizations of dynamic multi-exit algorithms may also be possible and are another direction for future work.

\bibliographystyle{IEEEtran}
\bibliography{name.bib}

\end{document}